\documentclass[prl,twocolumn,showpacs,preprintnumbers,amsmath,amssymb,groupedaddress,superscriptaddress]{revtex4}

\usepackage{graphicx}
\usepackage{dcolumn}
\usepackage{bm}
\usepackage{psfrag}
\usepackage{epsfig}
\usepackage{amsmath}
\usepackage{amssymb}
\usepackage{color}
\usepackage{fancyheadings}
\usepackage{mathbbol}

\newcommand{\nn}{\nonumber}
\newcommand{\bra}{\langle}
\newcommand{\ve}{\vert}
\newcommand{\ket}{\rangle}

\begin{document}


\title{Existence of Temperature on the Nanoscale}

\author{Michael Hartmann} 
\email{michael.hartmann@dlr.de}
\affiliation{Institute of Thechnical Physics, DLR Stuttgart, 70569 Stuttgart, Germany}
\affiliation{Institute of Theoretical Physics I, University of Stuttgart, 70550 Stuttgart, Germany}
\author{G\"unter Mahler}
\affiliation{Institute of Theoretical Physics I, University of Stuttgart, 70550 Stuttgart, Germany}
\author{Ortwin Hess}
\affiliation{Advanced Technology Institute, University of Surrey, Guildford GU2 7XH, United Kingdom}

\date{\today}

\begin{abstract}
We consider a regular chain of quantum particles with nearest 
neighbour interactions in a canonical state with temperature $T$. 
We analyse the conditions under which the state factors into a 
product of canonical density matrices with respect to groups of $n$ 
particles each and under which these groups have the same temperature $T$. 
In quantum mechanics the minimum group size $n_{min}$ depends on 
the temperature $T$, contrary to the classical case. We apply our 
analysis to a harmonic chain and find that $n_{min} = const.$ for 
temperatures above the Debye temperature and $n_{min} \propto T^{-3}$ below.
\end{abstract}

\pacs{05.30.-d, 05.70.Ce, 65.80.+n, 65.40.-b}
\maketitle

%
Recent progress in the synthesis and processing of materials with structures on
nanometer length scales calls for better understanding of thermal properties of
nanoscale devices, individual nanostractures and nanostructured materials.
Experimental techniques have improved to such an extent that the measurement of thermodynamic
quantities like temperature with a spatial resolution on the nanometer scale seems within reach
\cite{Gao2002,Pothier1997,Aumentado2001,Cahill2003}.
These techniques have already been applied for a new type of scanning microscopy,
using a temperature sensor \cite{Williams1986} that shows resolutions below 100nm.

To provide a basis for the interpretation of present day and future experiments in this field,
it is indispensable to clarify the applicability of the concepts of thermodynamics
to systems on small length scales. In this context, one question appears to be
particularly important and interesting: Can temperature be meaningfully defined on nanometer length scales?

The standard procedure to show that the thermodynamical limit and therefore temperature exists
are based on the idea that, as the spatial extension increases,
the surface of a region in space grows slower than its volume \cite{Ruelle1969}.
If the coupling potential is short-ranged enough, the interactions
between one region and another become negligible in the limit of infinite size.

However, the full scaling behavior of these interactions with respect to the size of the
parts has, to our knowledge, not been studied in detail yet \cite{Hill2001,Allahverdyan2002,Hartmann2003a}.

For standard applications of thermodynamics this might not be very important since the
number of particles is typically so large that deviations from infinite systems may safely be neglected.
Nevertheless, these differences could become significant as the considered systems approach nanoscopic
scales.
Here it is of special interest to determine the ``grain-size'' needed
to ensure the existence of local temperature.

Since a quantum description becomes imperative at nanoscopic scales,
the following approach appears to be reasonable:
Consider a large quantum system, brought into a thermal state via interaction with its environment,
divide this system into subgroups and analyse for what
subgroup-size the concept of temperature is applicable.

We adopt here the convention, that a local temperature exists, if the considered part of the system is in a
canonical state, i.e. the distribution is an exponentially decaying function of energy characterised by one
single parameter. This implies that there is a one to one mapping between temperature and
the expectation values of those observables, by which temperature is usually measured.
After proper calibration, such measurements thus all yield the same temperature,
contrary to distributions with several parameters.
If the distribution were not exponentially decaying, physical quantities like energy
would not have ``sharp'' values. 

Our results are thus obtained from calculations with canonical ensembles. Approaches based
on different ensembles (e.g. microcanonical ones), which are beeing discussed \cite{Gross2001},
may yield significantly different results.

In this paper we study a chain of identical particles with next neighbour interactions
which decomposes into $N_G$ identical subgroups of $n$ particles 
each.
Assuming that the total system is in a thermal (canonical) state, we derive conditions on the
interaction strength between the groups and the global temperature which ensure that the state of each group
is approximately canonical.
Finally, we apply these conditions to a harmonic chain and calculate the minimal group size $n_{min}$,
which is found to depend on the global temperature and
the desired accuracy of the local thermodynamical description.
%

We start by defining the Hamiltonian of our chain in the form,
\begin{equation}\label{hamil}
H = \sum_{i} H_i \enspace \enspace \textrm{with} \enspace \enspace
H_i = \frac{p_i^2}{2 m} + U(q_i) + V(q_i,q_{i+1}),
\end{equation}
where the index $i$ labels the particles of mass $m$ and $q_i$ and $p_i$ are their positions and momenta,
respectively.
We assume periodic boundary conditions.

We now form $N_G$ groups of $n$ subsystems each\linebreak
(index $i \rightarrow (\mu-1) n + j; \: \mu = 1, \dots, N_G; \: j = 1, \dots, n$)
and split this Hamiltonian into two parts,
\begin{equation}
\label{hsep}
H = H_0 + I,
\end{equation}
where $H_0$ is the sum of the Hamiltonians of the isolated groups
and $I$ contains the interaction terms of each group with its neighbour group:
\begin{eqnarray}\label{isogroups}
H_0 & = & \sum_{\mu=1}^{N_G} \left( \mathcal{H}_{\mu} - V(q_{\mu n}, q_{\mu n + 1}) \right) \enspace
; \enspace \mathcal{H}_{\mu} = \sum_{j=1}^n H_{n (\mu - 1) + j}\nn \\
I & = & \sum_{\mu=1}^{N_G} V(q_{\mu n}, q_{\mu n + 1}).
\end{eqnarray}
We label the eigenstates of the total Hamiltonian $H$
and their energies with the greek indices $(\varphi, \psi)$ and eigenstates and energies
of the group Hamiltonian  $H_0$ with latin indices $(a, b)$,
\begin{equation}
\label{prodstate}
H \: \ve \varphi \ket = E_{\varphi} \: \ve \varphi \ket \enspace \enspace \textrm{and} \enspace \enspace
H_0 \: \ve a \ket = E_a \: \ve a \ket.
\end{equation}
Here, the states $\ve a \ket$ are products of group eigenstates, with $E_a = \sum_{\mu=1}^{N_G} E_{\mu}$,
where $E_{\mu}$ is the energy of one subgroup only,
$\left( \mathcal{H}_{\mu} - V(q_{\mu n}, q_{\mu n + 1}) \right) \ve a \ket = E_{\mu} \ve a \ket$.

We assume that the total system is in a thermal state with a density matrix, which reads
\begin{equation}
\label{candens}
\bra \varphi \ve \hat \rho \ve \psi \ket = \frac{e^{- \beta E_{\varphi}}}{Z} \: \delta_{\varphi \psi}
\end{equation}
in the eigenbasis of $H$. Here, $Z$ is the partition sum and $\beta = (k_B T)^{-1}$
the inverse temperature.
Transforming the density matrix (\ref{candens}) into the eigenbasis of $H_0$ we obtain
\begin{equation}
\label{newrho}
\bra a \ve \hat \rho \ve a \ket =
\int_{E_0}^{\infty} w_a (E) \: \frac{e^{- \beta E}}{Z} \: dE
\end{equation}
for the diagonal elements. Here, the sum over all states $\ve a \ket$ has been replaced
by an integral over the energy. The conditional probabilities $w_a (E)$ are given by\linebreak
$w_a (E_{\varphi}) = \eta(E_{\varphi}) \, \sum_{\{H \ve \varphi \ket = E_{\varphi} \ve \varphi \ket\}}
\ve \bra a \ve \varphi \ket \ve^2$ with $\eta(E_{\varphi})$
denoting the density of energy levels and the sum running over all states $\ve \varphi \ket$ with
$H \ve \varphi \ket = E_{\varphi} \ve \varphi \ket$.
To compute the integral of equation (\ref{newrho}) we need to know the distribution of the
conditional probabilities $w_a (E)$.

In the limit of infinite number of groups $N_G$, 
$w_a (E_{\varphi})$ converges to a Gaussian normal distribution,
\begin{equation}
\label{gaussian}
\lim_{N_G \to \infty} w_a (E) = \frac{1}{\sqrt{2 \pi} \sigma_a}
\exp \left(- \frac{\left(E - E_a - \varepsilon_a \right)^2}{2 \, \sigma_a^2} \right),
\end{equation}
where the quantities $\varepsilon_a$ and $\sigma_a$ are defined by 
\begin{eqnarray}
\varepsilon_a \equiv \bra a \ve H \ve a \ket - \bra a \ve H_0 \ve a \ket, \\
\sigma_a^2 \equiv \bra a \ve H^2 \ve a \ket - \bra a \ve H \ve a \ket^2.
\end{eqnarray}
Note that $\varepsilon_a$ has a classical counterpart while $\sigma_a$
is purely quantum mechanical. It appears because the commutator $[H,H_0]$ is nonzero, and
the distribution $w_a(E)$ therefore has nonzero width. 

The rigorous proof of equation (\ref{gaussian}) is given in \cite{Hartmann2003} and is based on two
assumptions:
The energy of each group $\mathcal{H}_{\mu}$ as defined in equation (\ref{isogroups}) has to be bounded, i. e.
\begin{equation}
\label{bounded}
\bra \chi \ve \mathcal{H}_{\mu} \ve \chi \ket \le C
\end{equation}
for all normalised states $\ve \chi \ket$ and some arbitrary constant $C$, and
\begin{equation}
\label{vacuumfluc}
\sigma_a^2 \ge N_G \, C'
\end{equation}
for some constant $C' > 0$.

If these conditions are met, equation (\ref{newrho}) can be computed for $N_G  \gg 1$:
\begin{equation}
\label{newrho2}
\bra a \ve \hat \rho \ve a \ket =
\frac{\exp \left(- \beta y_a + \frac{1}{2} \beta^2 \sigma_a^2 \right)}{2 \: Z}  \: \,
\textrm{erfc} \left( \frac{E_0 - y_a + \beta \sigma_a^2}{\sqrt{2} \, \sigma_a} \right),
\end{equation}
where $y_a = E_{a} + \varepsilon_a$ and $\textrm{erfc} (x)$ is the conjugate Gaussian error function.
The second term appears because the energy is bounded from below and the
integration extends from $E_0$ to $\infty$.

If the argument of the error function divided by $\sqrt{N_G}$ is finite
(different from zero), the conjugate error function can be
approximated by its asymptotic expansion for $N_G \gg 1$ \cite{Abramowitz1970}:
$\textrm{erfc}(x) \approx \exp \left(- x^2 \right) / \sqrt{\pi} \, x$ for $x \rightarrow \infty $
and $\textrm{erfc}(x) \approx 2 + \exp \left(- x^2 \right) / \sqrt{\pi} \, x $ for $x \rightarrow - \infty$.

The off diagonal elements $\bra a \ve \hat \rho \ve b \ket$ vanish for\linebreak
$\ve E_a - E_b \ve > \sigma_a + \sigma_b$ because the overlap of the two distributions of conditional
probabilities becomes negligible. For $\ve E_a - E_b \ve < \sigma_a + \sigma_b$, the transformation
involves an integral over frequencies and thus these terms are significantly smaller than
the entries on the diagonal.

We now test under what conditions the density matrix $\hat \rho$ may be approximated by a product
of canonical density matrices with temperature $\beta_{loc}$ for each subgroup\linebreak
$\mu = 1, 2, \dots, N_G$. If we assume periodic boundary conditions,
all reduced density matrices are equal and their product
is of the form $\bra a \ve \hat \rho \ve a \ket \propto \exp(- \beta_{loc} E_a)$.
Since the trace of a matrix is invariant under basis transformations, it is sufficient to verify
the correct energy dependence of the density matrix in the product basis. Thus the logarithm of the rhs of
equation (\ref{newrho2}) must be a linear function of $E_a$ plus a possible constant.
It follows from the asymptotic expansion of the conjugate error function, that this can only be the
case for
\begin{equation}
\label{cond1}
\frac{1}{\sqrt{N_G}} \, \frac{E_{a} + \varepsilon_a - E_0 - \beta \sigma_a^2}{\sqrt{2} \,\sigma_a} > 0,
\end{equation}
meaning the lhs should have a finite positive value, and
\begin{equation}
\label{cond2}
- \varepsilon_a + \frac{\beta}{2} \sigma_a^2 \approx c_1 E_a + c_2,
\end{equation}
where $c_1$ and $c_2$ are constants.

If $| c_1 | \ll 1$ temperature becomes intensive:
\begin{equation}
\label{intensivity}
\beta_{loc} = \beta.
\end{equation}
Furthermore, to ensure, that the density matrix of each subgroup $\mu$ is approximately canonical,
it must hold that
\begin{equation} 
\label{subcondition}
- \frac{\varepsilon_{\mu-1} + \varepsilon_{\mu}}{2} + \frac{\beta}{4} \,
\left(\sigma_{{\mu}-1}^2 + \sigma_{\mu}^2 \right) + \frac{\beta}{6} \,\tilde{\sigma}_{\mu}^2 \approx
c_1 E_{\mu} + c_2, 
\end{equation}
where $\varepsilon_{\mu} = \bra a \ve V(q_{\mu n}, q_{\mu n + 1}) \ve a \ket$,
$\sigma_{\mu}^2 = \bra a \ve \mathcal{H}_{\mu}^2 \ve a \ket -
\bra a \ve \mathcal{H}_{\mu} \ve a \ket^2$ and
$\tilde{\sigma}_{\mu}^2 =
\sum_{\nu = \mu-1}^{\mu+1} \bra a \ve \mathcal{H}_{\nu-1} \mathcal{H}_{\nu} +
\mathcal{H}_{\nu} \mathcal{H}_{\nu-1} \ve a \ket -
2 \bra a \ve \mathcal{H}_{\nu-1} \ve a \ket \bra a \ve \mathcal{H}_{\nu} \ve a \ket$.

For a model obeying equations (\ref{bounded}) and (\ref{vacuumfluc}), the two conditions
(\ref{cond1}) and (\ref{subcondition}), which constitute the general result of this letter,
must both be satisfied. These fundamental criteria will now be applied to a concrete example.
%
%
%

Thermal properties of insulating solids are quite successfully described by harmonic lattice models. 
We therefore consider as an example a harmonic chain of $N_G \cdot n$ sites with periodic
boundary conditions. In this case, the potentials read
\begin{equation}
U(q_{i}) = \frac{m}{2} \, \omega_0^2 \, q_{i}^2 \enspace \enspace \textrm{and} \enspace \enspace
V(q_{i},q_{i+1}) = - m \, \omega_0^2 \, q_{i} \, q_{i+1},
\end{equation}
where $q_{i}$ is the displacement of the particle at site $i$ from its equilibrium
position $i \cdot a_0$ with $a_0$ being the distance between neighbouring particles at equilibrium.
We devide the chain into $N_G$ groups of $n$ particles each and thus get
a partition of the type considered above.
The Hamiltonian of one group is diagonalised by a Fourier transformation and reads in terms of creation
and anihilation operators $a_{k}^{\dagger}$ and $a_{k}$ \cite{Rieder1967},
\begin{equation}
\mathcal{H}_{\mu} - V(q_{\mu n}, q_{\mu n + 1})
= \sum_k \omega_k \left(a_{k}^{\dagger} a_{k} + \frac{1}{2} \right)
\end{equation}
with $k = \pi l / (a_0 \, (n+1))$ $(l = 1, 2, \dots, n)$ and\linebreak
$\omega^2_{k} = 4 \omega_0^2 \sin^2(k a / 2)$. We chose $\hbar = 1$.

We first check condition (\ref{vacuumfluc}). 
Expressing the group interaction $V(q_{\mu n}, q_{\mu n + 1})$ also in terms of
$a_{k}^{\dagger}$ and $a_{k}$,
one sees that $\tilde{\sigma}_{\mu} = 0$ for all $\mu$
and therefore $\sigma_a^2 = \sum_{\mu=1}^{N_G} \sigma_{\mu}^2$,
where $\sigma_{\mu}$ has a minimum value since $[a_{k}, a_{p}^{\dagger}] = \delta_{kp}$,\linebreak
($\sigma_{\mu}^2 \gtrsim 4 \omega_0 n / 3 \pi$).
$\sigma^2_a$ thus fulfills condition (\ref{vacuumfluc}).

To satisfy condition (\ref{bounded}) we only consider states where the energy of every group,
including the interactions with its neighbours, is bounded. Thus, our considerations do not apply
to product states $\ve a \ket$, for which all the energy was located in only one group or only a small
number of groups. For $N_G \gg 1$, the number of such states is vanishingly small compared
to the number of all product states.

To analyse conditions (\ref{cond1}) and (\ref{subcondition}), we make use of the continuum or
Debye approximation \cite{Kittel1983}, requiring $n \gg 1$ and $a_0 \ll l$, where $l = n \, a_0$,
the length of the chain, is finite.
In this case we have $\omega_k \approx v \, k$ with the constant velocity of
sound $v = \omega_0 \, a_0$ and $\sigma_{\mu}^2 = 4 \, n^{-2} \, E_{\mu} E_{\mu+1}$.

It is sufficient to fulfill conditions (\ref{cond1}) and (\ref{subcondition})
for some adequate energy range $E_{min} \le E_{\mu} \le E_{max}$.
We therefore need to estimate $E_{min}$ and $E_{max}$.
The total chain is in a thermal state. Let $\overline{E}$ be the expectation value of its energy without the
ground state energy,
\begin{equation}
\label{intenergy}
\overline{E} = N_G \, n \, k_B \Theta \left( \frac{T}{\Theta} \right)^2
\int_0^{\Theta / T} \frac{x}{e^x - 1} \, dx.
\end{equation}
Here, $\Theta$ is the Debye temperature, which is a characteristic constant
of the material chosen and can be found tabulated \cite{Kittel1983}. We then choose
\begin{equation}
\label{e_range}
\frac{1}{\alpha} \frac{\overline{E}}{N_G} + \frac{E_0}{N_G} \le E_{\mu}
\le \alpha \frac{\overline{E}}{N_G} + \frac{E_0}{N_G}
\end{equation}
with $\alpha \gg 1$. The product of the strongly growing density of states and the
exponentially decaying occupation probability forms a sharp peak around $\overline{E}$.
The above energy range has been chosen to be centered at this peak and to be significantly larger than its
width.

We now analyse conditions (\ref{cond1}) and (\ref{subcondition}) for the harmonic chain in Debye approximation.
In this case, condition (\ref{cond1}) translates into
\begin{equation}
\label{harmcond1}
\frac{1}{\sqrt{N_G}} \, \frac{\sum_{\mu=1}^{N_G} E_{\mu} - E_0 -
4 \, \beta \, n^{-2} \, \sum_{\mu=1}^{N_G} E_{\mu} E_{\mu+1}}
{\sqrt{2} \, \sqrt{4 \, n^{-2} \, \sum_{\mu=1}^{N_G} E_{\mu} E_{\mu+1}}} > 0,
\end{equation}
where $\varepsilon_{\mu} = 0$, $\tilde{\sigma}_{\mu} = 0$, $n+1 \approx n$ has been used.
For $\overline{E} > E_0$, condition (\ref{subcondition}) turns out to be stronger and (\ref{harmcond1})
becomes irrelevant.
For $\overline{E} < E_0$, (\ref{harmcond1}) is hardest to satisfy if all $E_{\mu}$ are equal
to the lower boundary of the energy range (\ref{e_range}). For $n$ one thus gets
\begin{equation}
\label{harmcond2}
n > 4 \: \frac{\Theta}{T} \: \frac{\alpha}{\overline{e}} \, \left(\frac{\overline{e}}{\alpha} + \frac{1}{4}
\right)^2,
\end{equation}
where $\overline{e} = \overline{E} / (n N_G k_B \Theta)$ and we have used\linebreak
$E_0 = n N_G k_B \Theta / 4$.

To investigate condition (\ref{subcondition}) we take the derivative with respect to $E_{\mu}$ on both sides,
\begin{equation}
\label{subcondition2}
\frac{\beta}{n^2} \left( E_{\mu-1} + E_{\mu+1} - 2 \, \frac{E_0}{N_G} \right) +
\frac{2 \beta}{n^2} \frac{E_0}{N_G} \approx c_1
\end{equation}
where we have seperated the energy dependent and the constant part on the lhs.
(\ref{subcondition2}) is satisfied if the energy dependent part, the first term, is much smaller than one.
Taking $E_{\mu-1}$ and $E_{\mu+1}$ equal to the upper bound in equation (\ref{e_range}), this yields
\begin{equation}
\label{subcondition3}
n > \frac{2 \alpha}{\delta} \: \frac{\Theta}{T} \: \overline{e},
\end{equation}
where the ``accuracy'' parameter $\delta \ll 1$ quantifies the value of the energy dependent part in
(\ref{subcondition2}).

Since, by virtue of equations (\ref{harmcond2}) and (\ref{subcondition3}),
the constant part in the lhs of (\ref{subcondition2}) satisfies
$(2 \beta / n^2) (E_0 / N_G) < \sqrt{\delta} / (\sqrt{2} \alpha) - \delta / \alpha^2 \ll 1$,
temperature is intensive.

\begin{figure}[t]
\psfrag{-4.1}{\small \raisebox{-0.1cm}{$10^{-4}$}}
\psfrag{-3.1}{\small \raisebox{-0.1cm}{$10^{-3}$}}
\psfrag{-2.1}{\small \raisebox{-0.1cm}{$10^{-2}$}}
\psfrag{-1.1}{\small \raisebox{-0.1cm}{$10^{-1}$}}
\psfrag{1.1}{\small \raisebox{-0.1cm}{$10^{1}$}}
\psfrag{2.1}{\small \raisebox{-0.1cm}{$10^{2}$}}
\psfrag{3.1}{\small \raisebox{-0.1cm}{$10^{3}$}}
\psfrag{4.1}{\small \raisebox{-0.1cm}{$10^{4}$}}
\psfrag{1}{}
\psfrag{2}{\small \hspace{+0.2cm} $10^{2}$}
\psfrag{3}{}
\psfrag{4}{\small \hspace{+0.2cm} $10^{4}$}
\psfrag{5}{}
\psfrag{6}{\small \hspace{+0.2cm} $10^{6}$}
\psfrag{7}{}
\psfrag{8}{\small \hspace{+0.2cm} $10^{8}$}
\psfrag{n}{\raisebox{0.1cm}{$n_{min}$}}
\psfrag{c1}{$\: T / \Theta$}
\epsfig{file=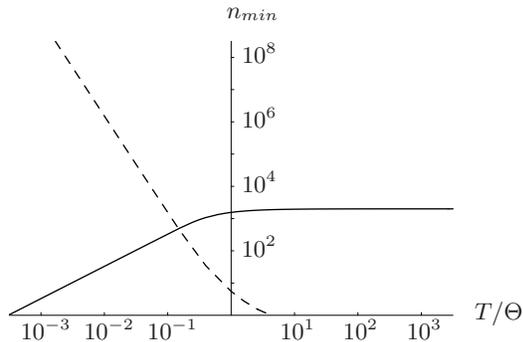,width=7cm}
\caption{Log-log-plot of $n_{min}$ from eq. (\ref{harmcond2}) (dashed line) and
$n_{min}$ from eq. (\ref{subcondition3}) (solid line) for $\alpha = 10$ and $\delta = 0.01$
as a function of $T / \Theta$ for a harmonic chain. $\delta$ and $\alpha$ are defined in equations
(\ref{subcondition3}) and (\ref{e_range}), respectively.}
\label{temp}
\end{figure}

Inserting equation (\ref{intenergy}) into equation (\ref{harmcond2}) and (\ref{subcondition3})
one can now calculate the minimal $n$ for given $\delta, \alpha,\Theta$ and $T$.
Figure \ref{temp} shows $n_{min}$ for $\alpha = 10$ and $\delta = 0.01$ given by criterion
(\ref{harmcond2}) and (\ref{subcondition3}), respectively, as a function of $T / \Theta$.
%
%
%

For high (low) temperatures $n_{min}$ can thus be estimated by
\begin{equation}
\label{high_low}
n_{min} \approx \left\{
\begin{array}{lcr}
2 \, \alpha / \delta & \textrm{for} & T > \Theta\\
\left( 3 \alpha / 2 \pi^2 \right) \, \left( \Theta / T \right)^3 & \textrm{for} & T < \Theta
\end{array}
\right.
\end{equation}
The characteristics of figure \ref{temp} are determined by the energy dependence of $\sigma_a$.
The low temperature behavior, in particular, is related to the fact that $\sigma_a$ has a nonzero
minimal value. This fact does not only appear in the harmonic chain but is a general feature
of interacting quantum particles. The commutator $[H,H_0]$ is nonzero and the ground state of the total
system is energetically lower than the lowest product state, therefore $\sigma_a$ is nonzero, even
at zero temperature \cite{Allahverdyan2002,Wang2002
}.

Finally, the minimal length scale on which temperature exists according to our approach is given by
\begin{equation}
\label{length}
l_{min} = n_{min} \, a_0,
\end{equation}
where $a_0$ is the lattice constant. Temperature measurements
with a higher resolution should no longer be interpreted in a standard way.

Let us conclude with some numerical estimates: Chosing the ``accuracy parameters''
to be $\alpha = 10$ and\linebreak $\delta = 0.01$,
we get for hot iron ($T \gg \Theta \approx 470 \,$K, $a_0 \approx 2.5 \,${\AA})
$l_{min} \approx 50 \,\mu$m, while
for carbon ($\Theta \approx 2230 \,$K, $a_0 \approx 1.5 \,${\AA}) at room temperature ($270 \,$K)
$l_{min} \approx 10 \,\mu$m. The coarse-graining will experimentally be most relevant at very low
temperatures,
where $l_{min}$ may even become macroscopic. A pertinent example is silicon
($\Theta \approx 645 \,$K, $a_0 \approx 2.4 \,${\AA}),
which has $l_{min} \approx 10 \,$cm at $T \approx 1 \,$K
(again with $\alpha = 10$ and $\delta = 0.01$).

In summary we have considered a linear chain of particles interacting with their nearest neighbours.
We have partitioned the chain into identical groups of $n$ adjoining particles each.
Taking the number of such groups to be very large and assuming the total
system to be in a thermal state with temperature $T$ we have found conditions
(equations (\ref{cond1}) and (\ref{subcondition})), which ensure that each group is
approximately in a thermal state with the same temperature $T$, implying intensivity.
In the quantum regime, these conditions depend on the temperature $T$, contrary to the classical case.

We have then applied our general method to a harmonic chain. Equations (\ref{high_low}) and (\ref{length})
provide a first estimate of the minimal length scale on which intensive
temperatures exist. This length scale should also constrain the way one can meaningfully
define temperature profiles in non-equilibrium scenarios \cite{Michel2003}.

We thank J.\ Gemmer, M.\ Michel, H.\ Schmidt, M.\ Stollsteimer, F.\ Tonner, M.\ Henrich and C.\ Kostoglou
for fruitful discussions.

\end{document}